\documentclass[apjl]{emulateapj}

\usepackage{amsmath}
\usepackage{graphics}
\shorttitle{Analytical Calculation of Non-linear Matter Power Spectrum}
\shortauthors{JEONG $\&$ KOMATSU}
\begin{document}
\title{%
  Perturbation Theory Reloaded: Analytical Calculation of
  Non-linearity in Baryonic Oscillations in the Real Space Matter 
  Power Spectrum
}%
\author{Donghui Jeong and Eiichiro Komatsu}
\affil{Department of Astronomy, University of Texas at Austin, \\
       1 University Station, C1400, Austin, TX, 78712}
\email{djeong@astro.as.utexas.edu}
\begin{abstract}
 We compare the non-linear matter power spectrum in real space calculated 
 analytically from 3rd-order perturbation theory 
 with $N$-body simulations at $1<z<6$. We find that the perturbation theory
  prediction
 agrees with the simulations to better than 1\%
 accuracy in the weakly non-linear regime where the dimensionless 
 power spectrum, 
 $\Delta^2(k)=k^3P(k)/2\pi^2$, which approximately gives variance of 
 matter density field at a given $k$, is less than 0.4. 
 While the baryonic acoustic oscillation features are preserved in the
 weakly non-linear regime at $z>1$, the shape of oscillations is 
 distorted from the linear theory prediction. Nevertheless, our results
 suggest that one can correct the distortion caused by non-linearity 
 {\it almost exactly}.
 We also find that perturbation theory, which does not contain any free
 parameters, provides a significantly better fit to the
 simulations than the conventional approaches based on empirical fitting
 functions to simulations. 
 The future work would include perturbation theory
 calculations of non-linearity in redshift space distortion and halo
 biasing in the weakly non-linear regime.
\end{abstract}
\section{Introduction}
Cosmological linear perturbation theory
has been remarkably successful in explaining the precision measurements
of temperature and polarization anisotropies of the cosmic
microwave background (CMB), most notably from the {\sl
Wilkinson Microwave Anisotropy Probe} ({\sl WMAP})
\cite{bennett/etal:2003}.
The CMB data, combined with linear theory, have enabled us to determine
many of the cosmological parameters to better than 10\%
accuracy \cite{spergel/etal:2006}.
As the CMB data improve, however, it has become increasingly clear that one has
to combine the CMB data with the other probes 
to break degeneracies between the parameters that
cannot be constrained very well by the CMB data alone. For example, the CMB alone cannot break
degeneracy between the equation of state of dark energy, $w$, and
matter density, $\Omega_m$ \cite{spergel/etal:2006}.

The large-scale structure (LSS) of the universe has also been known as
an excellent probe of cosmological fluctuations as well as cosmological
parameters, as proven successfully by the {\sl Two-degree
Field Galaxy Redshift Survey} ({\sl 2dFGRS}) \cite{cole/etal:2005} and 
the {\sl Sloan Digital Sky Survey} ({\sl SDSS}) \cite{tegmark/etal:2004,seljak/etal:2005}.
A joint analysis of the future CMB and LSS data is 
extremely powerful in constraining most of the cosmological parameters to 
better than a few percent accuracy \cite[e.g.,][]{takada/komatsu/futamase:2006}.
In particular, the LSS data would allow us to 
constrain ``additional'' parameters such as the mass of neutrinos
and the shape of the primordial power spectrum,
which would remain relatively poorly constrained by the CMB data alone.

The success of this approach depends on our ability to
predict the power spectrum of CMB and LSS from theory. Linear theory
provides adequate precision for CMB, as the amplitude of CMB anisotropy
is only $10^{-5}$; however, theory of LSS has not 
reached to the point where one can use LSS for precision
cosmology at the level similar to CMB. There is a larger 
degree of non-linearity in LSS. In order for the LSS data to be as
powerful as the CMB data, it is crucial that we can predict the LSS
power spectrum to 1\% accuracy, which is nearly one order of magnitude
better than the current precision. 

In principle, theory of LSS
may be developed using $N$-body simulations. 
This approach has been widely used in the literature. 
One method builds on the so-called HKLM formalism
\cite{hamilton/etal:1991}, which interpolates between the linear regime
on large scales and the stable clustering regime on small scales using a
fitting function to $N$-body simulations. The HKLM method was
further elaborated by \cite{peacock/dodds:1996}.
The other method builds on the so-called halo model 
\cite{scherrer/bertschinger:1991}, which was further elaborated 
by e.g., \cite{seljak:2000,smith/etal:2003}.
Both approaches are based on {\it empirical} methods, fitting
to $N$-body simulations mainly at $z\sim 0$. While
these predictions may be good to within 10\%, one should not
expect 1\% accuracy from these.
Also, these methods, in their current form, do not
allow for non-linearity in redshift space distortion in the weakly
non-linear regime, which limits their practical use for
the actual data analysis. 

We use an alternative approach based on cosmological perturbation
theory (PT). One can calculate the next-to-leading order correction to the
linear power spectrum by using 3rd-order PT
\cite{vishniac:1983,suto/sasaki:1991,makino/sasaki/suto:1992,jain/bertschinger:1994,scoccimarro/frieman:1996}. 
The advantage of PT is that it provides an {\it exact}
solution for the non-linear matter power spectrum 
as long as one applies it to the region in $k$ space where
perturbative expansion is valid. (We shall call this region in $k$ space the
``weakly non-linear regime''.) One still needs to use simulations to
find the maximum $k$ below which perturbation expansion is valid, 
which is one of the goals of this paper.

Cosmological PT, including non-linear corrections to
the power spectrum, was
actively investigated in 1990's \cite[][for a review]{bernardeau/etal:2002}.
In particular, a lot of efforts have been devoted into understanding the
non-linear power spectrum at $z\sim 0$. It was shown that
perturbation approach would not provide accurate descriptions of the
power spectrum at $z\sim 0$ due to too strong non-linearity. 
Our results are consistent with the previous work; however, we focus
on the power spectrum at $z>1$, where non-linearity is still
modest and thus PT should perform better. 

Our work is
motivated by recent proposals of high-$z$ galaxy survey projects such as 
the {\sl Cosmic Inflation Probe} ({\sl CIP}) \cite{melnick/etal:2005},
{\sl Hobby-Ebery Dark Energy Experiment} ({\sl HETDEX})
\cite{hill/etal:2004}, and 
{\sl Wide-field Fiber-fed Multi Object Spectrograph}
({\sl WFMOS}) survey \cite{glazebrook/etal:2005}, to mention
a few. The goal of  these missions is to measure the
power spectrum of high-$z$ galaxies to a few percent accuracy. 
These missions should be able to measure the baryonic
features in the power spectrum accurately.
On the other hand, it has been pointed out  that
non-linearity would distort the baryonic features
in a complex way so that it might be challenging to extract the
underlying baryonic features from the observed galaxy power spectrum
\cite{meiksin/white/peacock:1999,springel/etal:2005,white:2005,seo/eisenstein:2005}.
We show that, as far as non-linearity in the matter power spectrum in
real space is concerned, we can correct it almost exactly.

This paper is organized as follows. We briefly review the
3rd-order PT in \S~2, and describe our $N$-body
numerical simulations in \S~3. We compare the analytical predictions
with simulations in \S~4. We pay a particular attention to
non-linearity in the baryonic acoustic oscillations. We give discussion and
conclusions in \S~5. We test convergence of our results in Appendix~\ref{sec:app}.

\section{Non-linear Matter Power spectrum: 3rd-order Perturbation
 Theory}
We review 3rd-order PT
calculations of the next-to-leading order correction to the matter power
spectrum, following the pioneering work in the literature
\cite{vishniac:1983,fry:1984,goroff/etal:1986,suto/sasaki:1991,makino/sasaki/suto:1992,jain/bertschinger:1994,scoccimarro/frieman:1996}.
As the power spectrum, $P(k,\tau)$,
 is a quadratic quantity of the density field in Fourier space, 
$\tilde{\delta}_{\mathbf k}(\tau)$, 
\begin{equation}
 \langle \tilde{\delta}_{\mathbf k}(\tau)\tilde{\delta}^*_{{\mathbf
  k}'}(\tau)\rangle = (2\pi)^3P(k,\tau)
  \delta_D({\mathbf k}-{\mathbf k}'),
\end{equation}
the 3rd-order expansion in the density field is necessary for obtaining the
next-to-leading order correction to $P(k,\tau)$. 
We often use the ``dimensionless power spectrum'', $\Delta^2(k,\tau)$,
which represents the contribution to the variance of density field per
$\ln k$,
\begin{equation}
 \langle \delta^2({\mathbf x},\tau)\rangle
= \int \frac{dk}{k}~\Delta^2(k,\tau),
\label{eq:deltasq}
\end{equation}
where 
$\Delta^2(k,\tau)\equiv k^3P(k,\tau)/(2\pi^2)$.

We treat dark matter and baryons
as pressureless dust particles, as we
are interested in the scales much larger than the Jeans
length.
We also assume that peculiar velocity is much smaller than the
speed of light, which is always an excellent approximation, and that
the fluctuations we are interested in are deep inside the horizon; thus,
we treat the system as Newtonian.
The basic equations to solve are given by 
\begin{equation}
   \dot\delta 
   + \nabla\cdot[(1+\delta)\mathbf{v}]
   =0, \label{cont}                                     
\end{equation}
\begin{equation}
   \dot{\mathbf{v}} 
   + (\mathbf{v} \cdot\nabla) \mathbf{v} 
   = 
   -\frac{\dot{a}}{a} \mathbf{v} - \nabla\phi, \label{euler}
\end{equation}
\begin{equation}
   \nabla^2\phi 
   =
   4\pi G a^2 \bar{\rho} \delta, \label{poisson}
\end{equation}
where 
the dots denote $\partial/\partial\tau$
($\tau$ is the conformal time),
$\nabla$ denotes $\partial/\partial{\mathbf x}$
(${\mathbf x}$ is the comoving coordinate), 
${\mathbf
v}=d{\mathbf x}/d\tau$ is the peculiar velocity field, 
and $\phi$ is the peculiar gravitational potential field
from density fluctuations.
We assume that ${\mathbf v}$ is curl-free, which
motivates our using $\theta \equiv \nabla\cdot\mathbf{v}$, 
the velocity divergence field.
Using equation~(\ref{poisson}) and the Friedmann equation, 
we write the continuity equation [Eq.~(\ref{cont})] and 
the Euler equation [Eq.~(\ref{euler})] in Fourier space as
\begin{eqnarray}
\nonumber
& & 
\dot{\tilde{\delta}}_{\mathbf k}(\tau)  + \tilde{\theta}_{\mathbf k}(\tau)  \\
    &= &
      -\int \frac{d^3 k_1}{(2\pi)^3} \int d^3 k_2 
       \delta_D (\mathbf{k}_1+\mathbf{k}_2-\mathbf{k})
      \frac{\mathbf{k}\cdot\mathbf{k}_1}{k^2_1}
      \tilde{\delta}_{{\mathbf k}_2}(\tau) 
      \tilde{\theta}_{{\mathbf k}_1}(\tau), \\
\nonumber
& &    \dot{\tilde{\theta}}_{\mathbf k}(\tau)  + \frac{\dot{a}}{a}
   \tilde{\theta}_{\mathbf k}(\tau)  
+ \frac{3\dot{a}^2}{2a^2}\Omega_{\rm m}(\tau)\tilde{\delta}_{\mathbf k}(\tau)\\ 
\nonumber
    &= & 
      - \int \frac{d^3 k_1}{(2\pi)^3} \int d^3 k_2
        \delta_D(\mathbf{k}_1+\mathbf{k}_2-\mathbf{k}) 
      \frac
           {k^2 (\mathbf{k}_1\cdot\mathbf{k}_2)}
           {2 k_1^2 k_2^2} 
      \tilde{\theta}_{{\mathbf k}_1}(\tau)
      \tilde{\theta}_{{\mathbf k}_2}(\tau),\\
\end{eqnarray}
respectively. 

To proceed further, we assume that the universe is matter dominated,
$\Omega_{\rm m}(\tau)=1$ and $a(\tau)\propto \tau^2$. 
Of course, this assumption cannot be fully justified, as 
dark energy dominates the universe at low $z$. Nevertheless, it has been
shown that the next-to-leading order correction to
$P(k)$ is extremely insensitive to the underlying cosmology, if one uses
the correct growth factor for $\tilde{\delta}_{\mathbf k}(\tau)$
\cite{bernardeau/etal:2002}. Moreover, as we are primarily interested in
$z\geq 1$, where the universe is still matter dominated, accuracy of our
approximation is even better. (We quantify the error due to this
approximation below.)
To solve these coupled equations, we shall expand 
$\tilde{\delta}_{\mathbf k}(\tau)$ and $\tilde{\theta}_{\mathbf
k}(\tau)$ perturbatively using the $n$-th power of linear solution, 
$\delta_1(\mathbf{k})$, as a basis:
\begin{eqnarray}
\nonumber
 \tilde{\delta}(\mathbf{k},\tau) &=& \sum_{n=1}^{\infty} a^n (\tau)
   \int \frac{d^3 q_1}{(2\pi)^3} \cdots \frac{d^3 q_{n-1}}{(2\pi)^3}\\
\nonumber
   & &\times
\int d^3 q_n
        \delta_D (\sum_{i=1}^n \mathbf{q}_i - \mathbf{k})  \\
& &\times
        F_n (\mathbf{q}_1 ,\mathbf{q}_2 ,\cdots , \mathbf{q}_n)
        \delta_1(\mathbf{q}_1)\cdots \delta_1(\mathbf{q}_n),\\
\nonumber
 \tilde{\theta}(\mathbf{k},\tau) &=& -\sum_{n=1}^{\infty} \dot{a}(\tau)
a^{n-1}(\tau)
   \int \frac{d^3 q_1}{(2\pi)^3} \cdots \frac{d^3 q_{n-1}}{(2\pi)^3}  \\
\nonumber
   & &\times  \int d^3 q_n
\delta_D (\sum_{i=1}^n \mathbf{q}_i - \mathbf{k})    \\
& &\times
        G_n (\mathbf{q}_1 ,\mathbf{q}_2 ,\cdots , \mathbf{q}_n)
        \delta_1(\mathbf{q}_1)\cdots\delta_1(\mathbf{q}_n).
\end{eqnarray}
Here, the functions $F$ and $G$ as well as their recursion relations 
are given in \cite{jain/bertschinger:1994}.
As the linear density field, $\delta_1$, is a Gaussian random field, 
the ensemble average of odd powers of $\delta_1$ vanishes.
Therefore, the next-to-leading order correction to $P(k)$ is 
\begin{equation}
\label{eq:next-to-leading}
   P(k,\tau) = 
a^2(\tau) P_{11}(k) + a^4(\tau) [2P_{13}(k) + P_{22}(k)],
\end{equation}
where
\begin{equation}
\label{eq:P22}
P_{22}(k) = 
2 \int \frac{d^3 q}{(2\pi)^3} P_{11}(q) P_{11}(|\mathbf{k}-\mathbf{q}|)
            \left[F_2^{(s)}(\mathbf{q},\mathbf{k}-\mathbf{q})\right]^2,
\end{equation}
\begin{equation}
\label{eq:F2s}
  F_2^{(s)}(\mathbf{k}_1,\mathbf{k}_2)=
      \frac{5}{7}
    + \frac{2}{7}\frac{(\mathbf{k}_1\cdot\mathbf{k}_2)^2}{k_1^2k_2^2}
    + \frac{\mathbf{k}_1\cdot\mathbf{k}_2}{2}
      \left(
         \frac{1}{k_1^2}+\frac{1}{k_2^2}
       \right),
\end{equation}
\begin{eqnarray}
 \nonumber
   2P_{13}(k) &=&
      \frac{2\pi k^2}{252} P_{11}(k) \int_{0}^{\infty} \frac{dq}{(2\pi)^3}
      P_{11}(q)\\
\nonumber
& \times& \Biggl[100\frac{q^2}{k^2} -158 + 12\frac{k^2}{q^2} 
   -42 \frac{q^4}{k^4}\\
\label{eq:P13}
&+&\frac{3}{k^5q^3}(q^2-k^2)^3(2k^2+7q^2)\ln\left( \frac{k+q}{|k-q|} \right)
      \Biggl].
\end{eqnarray}
While $F_2^{(s)}({\mathbf k}_1,{\mathbf k}_2)$ should be 
modified for different cosmological models, the difference vanishes
when ${\mathbf k}_1\parallel {\mathbf k}_2$. The biggest
correction comes from the configurations with 
${\mathbf k}_1\perp {\mathbf k}_2$, for which 
$[F_2^{(s)}(\Lambda{\rm CDM})/F_2^{(s)}({\rm
EdS})]^2\simeq 1.006$ and $\lesssim 1.001$ 
at $z=0$ and $z\geq 1$, respectively.
Here, $F_2^{(s)}({\rm EdS})$ is given by equation~(\ref{eq:F2s}), while
$F_2^{(s)}(\Lambda{\rm CDM})$ contains corrections due to $\Omega_{\rm
m}\neq 1$ and $\Omega_\Lambda\neq 0$ 
\cite{matsubara:1995,scoccimarro/etal:1998},
and we used $\Omega_{\rm m}=0.27$ and $\Omega_\Lambda=0.73$ at present.
The information about different background cosmology 
is thus almost entirely encoded in the linear growth factor.
We extend the results obtained above to arbitrary cosmological models
by simply replacing $a(\tau)$ in equation~(\ref{eq:next-to-leading})
with an appropriate linear growth factor, $D(z)$,
\begin{equation}
   \label{eq:next-to-leading2}
   P(k,z) = 
D^2(z) P_{11}(k) + D^4(z) [2P_{13}(k) + P_{22}(k)].
\end{equation}
We shall use
equation~(\ref{eq:P22})--(\ref{eq:next-to-leading2})
to compute $P(k,z)$.

\section{$N$-body simulations and analysis method}
We use the TVD \citep{TVD} code to simulate the evolution of
$\delta({\mathbf x},\tau)$.
The TVD code uses the Particle-Mesh scheme for gravity,
and the Total-Variation-Diminishing (TVD) 
scheme for hydrodynamics, although
we do not use hydrodynamics in our calculations.
To increase the dynamic range of the derived power spectrum
and check for convergence of the results, we use four box sizes,
$L_{\rm box}=512$, 256, 128, and 64~$h^{-1}$~Mpc, 
with the same number of particles, $N=256^3$.
(We use $512^3$ meshes for doing FFT.)
We use the following cosmological parameters: 
$\Omega_{\rm m}=0.27$, $\Omega_{\rm b} = 0.043$, 
$\Omega_\Lambda = 0.73$, $h=0.7$, $\sigma_8 = 0.8$, and $n_{\rm s}=1$.
We output the simulation data at $z=6$, 5, 4, 3, 2
and 1 for 512, 256 and 128~$h^{-1}$~Mpc,
while only at $z=6$, 5, 4 and 3 for 64~$h^{-1}$~Mpc.

We suppress sampling variance of the estimated $P(k,z)$ by 
averaging $P(k,z)$ from 60, 60, 20, and 15 independent realizations of
 512, 256, 128, and 64~$h^{-1}$~Mpc simulations, respectively.
We calculate the density field on $512^3$ mesh points 
from the particle distribution by 
the Cloud-In-Cell (CIC) mass distribution scheme.
We then Fourier transform the density field and
average $|\delta_{\mathbf{k}}(\tau)|^2$ 
within $k-\Delta k/2\leq |{\mathbf k}|<k+\Delta k/2$
over the angle to estimate $P(k,z)$. Here, $\Delta k = 2\pi/L_{\rm box}$.
Finally, we correct the estimated $P(k)$ for loss of power due to 
the CIC pixelization effect
using the window function calculated from 100 realizations of 
random particle distributions.

\begin{figure}
\centering
\rotatebox{90}{%
  \includegraphics[width=6.5cm]{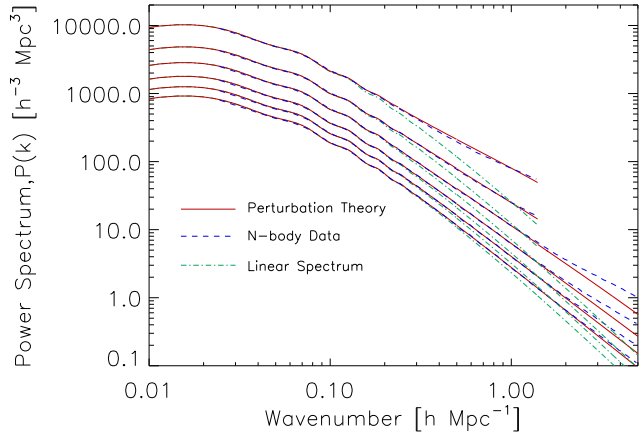}
}%
\caption{%
  Power spectrum at $z=1$, 2, 3, 4, 5 and 6 
 (from top to bottom), derived from $N$-body simulations (dashed lines),
 perturbation theory (solid lines), and linear theory
 (dot-dashed lines). We plot the simulation data from 512, 
 256, 128, and 64~$h^{-1}$~Mpc simulations
 at $k\leq 0.24~h~{\rm Mpc}^{-1}$, $0.24<k\leq 0.5~h~{\rm Mpc}^{-1}$,
$0.5<k\leq 1.4~h~{\rm Mpc}^{-1}$, and
$1.4<k\leq 5~h~{\rm Mpc}^{-1}$,
respectively. Note that 
we did not run 64~$h^{-1}$~Mpc simulations at $z=1$ or 2.
}%
\label{fig1}
\end{figure}

We use the {\sf COSMICS} package \cite{bertschinger:1995}
to calculate the linear transfer function (with {\sf linger}) 
and generate the input linear matter power spectrum and 
initial conditions (with {\sf grafic}).
We have increased the number of sampling points for the transfer function
in $k$ space from the default value of {\sf COSMICS}, 
as the default
sampling rate is too low to sample the baryonic acoustic oscillations
accurately. (The default rate resulted in an artificial numerical
smoothing of the oscillations.)
We locate initial particles on the regular grid (i.e., we do not
randomize the initial particle distribution),
and give each particle 
the initial velocity field using the Zel'dovich approximation.
This procedure  suppresses shot noise in the derived 
power spectrum, which arises from randomness of particle distribution.
We have checked this by comparing $P(k,z)$ from
the initial condition to the input linear spectrum.
However, some shot noise would arise as density fluctuations grow over time.
While it is difficult to calculate the magnitude of 
shot noise from the structure formation,
we estimate it by
comparing $P(k,z)$ from large-box simulations with that from 
small-box simulations. We do not find any evidence for shot
noise at $z\geq 1$; thus, we do not subtract shot noise from the 
estimated $P(k,z)$. To be conservative, 
we use 512, 256, 128, and 64~$h^{-1}$~Mpc simulations
to obtain $P(k,z)$ 
at $k\leq 0.24~h~{\rm Mpc}^{-1}$, $0.24<k\leq 0.5~h~{\rm Mpc}^{-1}$,
$0.5<k\leq 1.4~h~{\rm Mpc}^{-1}$, and
$1.4<k\leq 5~h~{\rm Mpc}^{-1}$, respectively, to avoid the residual
CIC pixelization effect and potential contaminations from unaccounted 
shot noise terms as well as artificial ``transients'' from 
initial conditions generated by the Zel'dovich approximation
\cite{crocce/etal:2006}. The initial redshifts are 
$z_{\rm initial}=27$, 34, 42, and 50
for 512, 256, 128, and 64~$h^{-1}$~Mpc simulations, respectively.
In Appendix~\ref{sec:app} we show more on the convergence test
(see Fig.~\ref{fig5}).

\section{Results}
Figure \ref{fig1} compares $P(k,z)$ at $z=1$, 2, 3, 4, 5 and 6 
(from top to bottom) 
from simulations (dashed lines), 
PT (solid lines), and linear theory
(dot-dashed lines). The PT predictions agree
with simulations so well that it is actually difficult to see 
the difference between PT and simulations in Figure~\ref{fig1}.
The simulations are significantly above the linear theory
predictions at high $k$.

\begin{figure}
\centering
\rotatebox{90}{%
  \includegraphics[width=6.5cm]{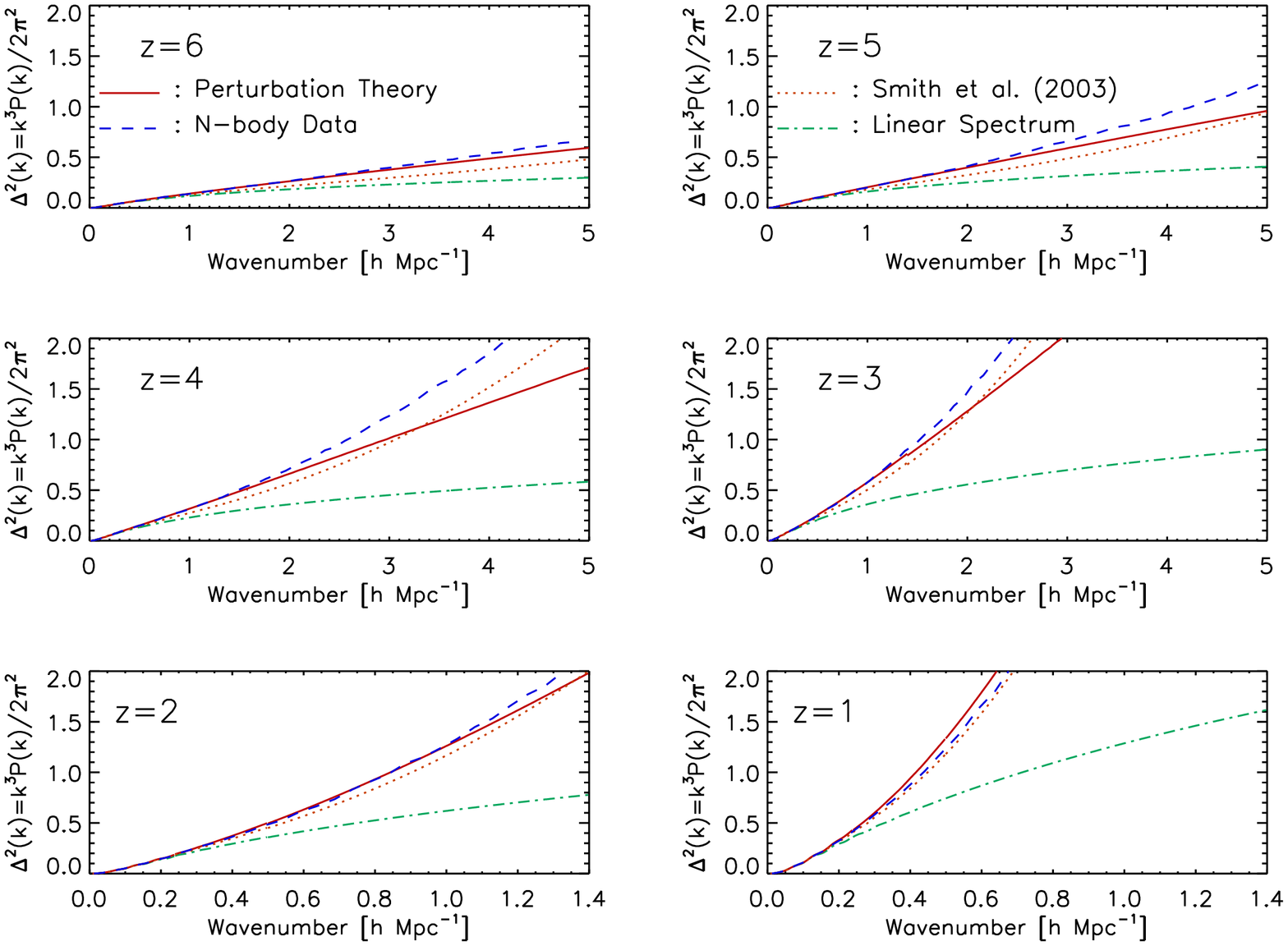}
}
\rotatebox{90}{%
  \includegraphics[width=6.5cm]{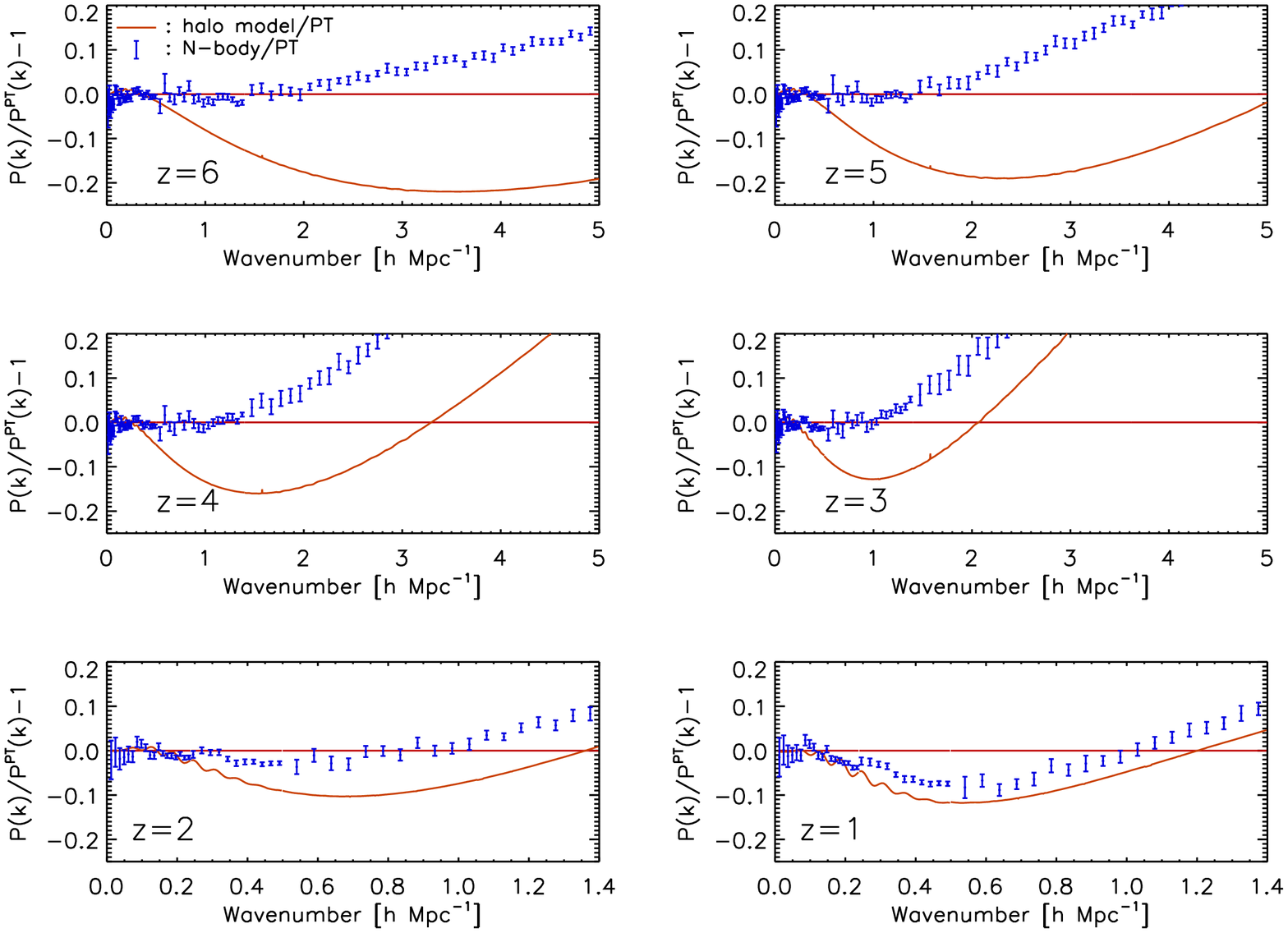}
}%
\caption{%
 ({\it Top}) Dimensionless power spectrum,  $\Delta^2(k)$.
  The solid and dashed lines show perturbation theory calculations and 
  $N$-body simulations, respectively. The dotted lines show
 the predictions from halo approach \cite{smith/etal:2003}. 
 The dot-dashed lines show the linear power spectrum. 
 ({\it Bottom}) Residuals. The errorbars show the $N$-body data divided
 by the perturbation theory predictions minus one, while the solid curves
 show the halo model calculations given in \cite{smith/etal:2003} divided by 
the perturbation theory predictions minus one.
 The perturbation theory predictions agree with simulations
 to better than 1\% accuracy for $\Delta^2(k)\lesssim 0.4$.
}%
\label{fig2}
\end{figure}

To facilitate the comparison better, we show $\Delta^2(k,z)$
[Eq.~(\ref{eq:deltasq})]
in Figure~\ref{fig2}.
We find that the PT predictions (thin solid lines) 
agree with simulations (thick solid lines) 
 to better than 1\% accuracy for $\Delta^2(k,z)\lesssim 0.4$.
On the other hand, the latest predictions from halo approach
\cite{smith/etal:2003} (dotted lines) perform significantly worse
then PT.
This result suggests that one must use PT
to model non-linearity in the weakly non-linear regime.

\begin{figure}
\centering
\rotatebox{90}{%
  \includegraphics[width=6.5cm]{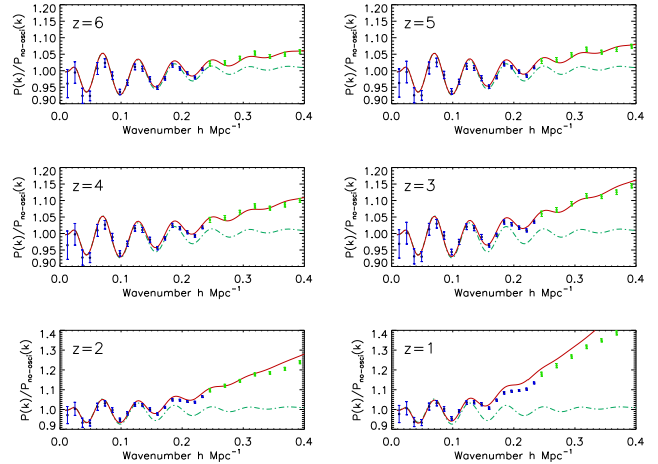}
}%
\caption{%
  Non-linearity in baryonic acoustic oscillations. All of the power spectra
 have been divided by a smooth power spectrum without baryonic
 oscillations from equation~(29) of \cite{eisenstein/hu:1998}.
 The errorbars show $N$-body simulations, while the solid lines 
 show perturbation theory calculations.
 The dot-dashed lines show the linear theory predictions.
 Perturbation theory describes non-linear distortion 
 on baryonic oscillations very accurately at $z>1$.
 Note that different redshift bins are not independent, as they have
 grown from the same initial conditions. The $N$-body data at
 $k<0.24$ and $k>0.24~h~{\rm Mpc}^{-1}$ are from
 512~and 256~$h^{-1}$~Mpc box simulations, respectively.
}%
\label{fig3}
\end{figure}

The baryonic features in the matter power spectrum
provide a powerful tool to
constrain the equation of state of dark energy. This method uses the
fact that the CMB angular power spectrum sets the physical acoustic scale, and
thus the features in the matter power spectrum seen on the sky and
in redshift space may be used as the
standard ruler, giving us the angular diameter distance out
to the galaxy distribution at a given survey redshift as well as $H(z)$
\cite{matsubara/szalay:2003,hu/haiman:2003,seo/eisenstein:2003,blake/glazebrook:2003}. 
In order for this method to be viable, however, it is crucial to
understand distortion on the baryonic acoustic oscillations caused by
non-linearity. This has been investigated so far mostly using direct
numerical simulations
\cite{meiksin/white/peacock:1999,springel/etal:2005,white:2005,seo/eisenstein:2005}.
\cite{meiksin/white/peacock:1999} also compared the PT prediction with
their $N$-body simulations at $z=0$, finding that PT was a poor fit.
This is because non-linearity at $z=0$ is too strong to model by PT.
Figure~\ref{fig3} shows 
that PT provides an accurate {\it analytical} account
of non-linear distortion at $z>1$: even at $z=1$, the 
third peak at $k\simeq 0.18~h~{\rm Mpc}^{-1}$ is modeled at a few
percent level.
At $z>2$, all the oscillatory features are modeled to
better than 1\% accuracy.
A slight deficit in power from $N$-body simulations at $k\sim 0.2~h~{\rm
Mpc}^{-1}$ relative to the perturbation theory predictions at $z=2$
may be due to artificial transient modes from the Zel'dovich
approximation used to generate initial conditions. One may eliminate
such an effect by either using a smaller box-size or a better initial
condition from the second-order Lagrangian perturbation theory
\cite{crocce/etal:2006}. As the power spectrum at $k>0.24~h~{\rm
Mpc}^{-1}$ from 256~$h^{-1}$~Mpc simulations at $z=2$ agrees with the
perturbation theory predictions very well, we conclude that this small
deficit in power at $k\sim 0.2~h~{\rm Mpc}^{-1}$ is a numerical effect,
most likely the transients in low-resolution simulations.  

\begin{figure}
\centering
\rotatebox{90}{%
  \includegraphics[width=6.5cm]{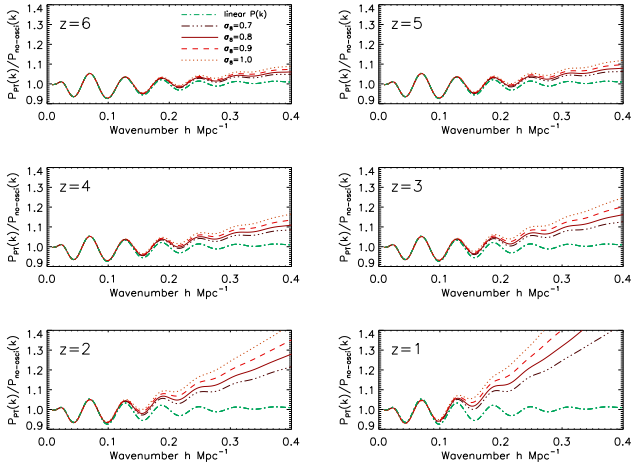}
}%
\caption{%
  Non-linearity and the amplitude of matter fluctuations, $\sigma_8$. 
In each panel the lines show the linear spectrum and non-linear
 spectrum with $\sigma_8=0.7$, 0.8, 0.9 and 1.0 from bottom to top. 
}%
\label{fig4}
\end{figure}

How do the predicted non-linear power spectra depend on 
the amplitude of matter fluctuations? 
As the non-linear contributions to the 
power spectrum are given by the linear spectrum squared, 
a non-linear to linear ratio grows in proportion to $\sigma_8^2$.
In Fig~\ref{fig4} we show how the non-linear contributions increase
as one increases $\sigma_8$ from 0.7 to 1.0. This figure may be useful
when one compares our results with the previous work that uses
different values of $\sigma_8$.

\section{Discussion and Conclusions}
The next-to-leading order correction to the matter power spectrum
calculated analytically
from 3rd-order PT provides an almost exact description
of the matter power spectrum in real space in the weakly
non-linear regime, where $\Delta^2(k)\lesssim 0.4$ (Fig.~\ref{fig2}).
The most important implications of our results for 
the planned high-$z$ galaxy surveys are that we can use PT
to calculate
(a) non-linearity in the baryonic acoustic oscillations (Fig.~\ref{fig3}), 
which should reduce systematics in constraining
dark energy properties, and (b) the matter power
spectrum up to much higher $k$ than that was accessible before, 
which should vastly increase our ability to measure
the shape of the primordial power spectrum as
well as the mass of neutrinos \cite{takada/komatsu/futamase:2006}. 
Of course, these surveys measure the {\it galaxy}
power spectrum in {\it redshift} space; thus,
the future work should include PT calculations of non-linearity in 
(a) redshift space distortion \cite{scoccimarro:2004}, and (b) halo
biasing \cite{fry/gaztanaga:1993,heavens/matarrese/verde:1998},
as well as an extensive comparison with numerical simulations.
PT also allows one to calculate the higher-order statistics such as
the bispectrum, which has been shown to be 
a powerful tool to check for systematics in our understanding of 
non-linear galaxy bias 
\cite{matarrese/verde/heavens:1997,verde/heavens/matarrese:1998}.
We should therefore ``reload'' cosmological perturbation theory
 and make a serious assessment of its validity 
in light of the planned high-$z$ galaxy surveys constraining
properties of dark energy, inflation, and neutrinos.\\

We would like to thank D. Ryu for letting us use his TVD code,
and K. Gebhardt, Y. Suto and M. Takada for comments.
D.J. would like to thank K. Ahn for his help on the TVD code.
E.K. acknowledges support from an Alfred P. Sloan Fellowship.
The simulations were carried out at the Texas Advanced Computing
Center (TACC).

\appendix
\section{Convergence Test}\label{sec:app}
To test convergence of the power spectra derived from
simulations and determine the valid range in wavenumber
from each simulation box, we have run $N$-body simulations
with four different box sizes,
$L_{\rm box}=512$, 256, 128, and 64~$h^{-1}$~Mpc, 
with the same number of particles, $N=256^3$.
The initial redshifts are 
$z_{\rm initial}=27$, 34, 42, and 50
for 512, 256, 128, and 64~$h^{-1}$~Mpc simulations, respectively.

\begin{figure}
\centering
\rotatebox{90}{%
  \includegraphics[width=6.5cm]{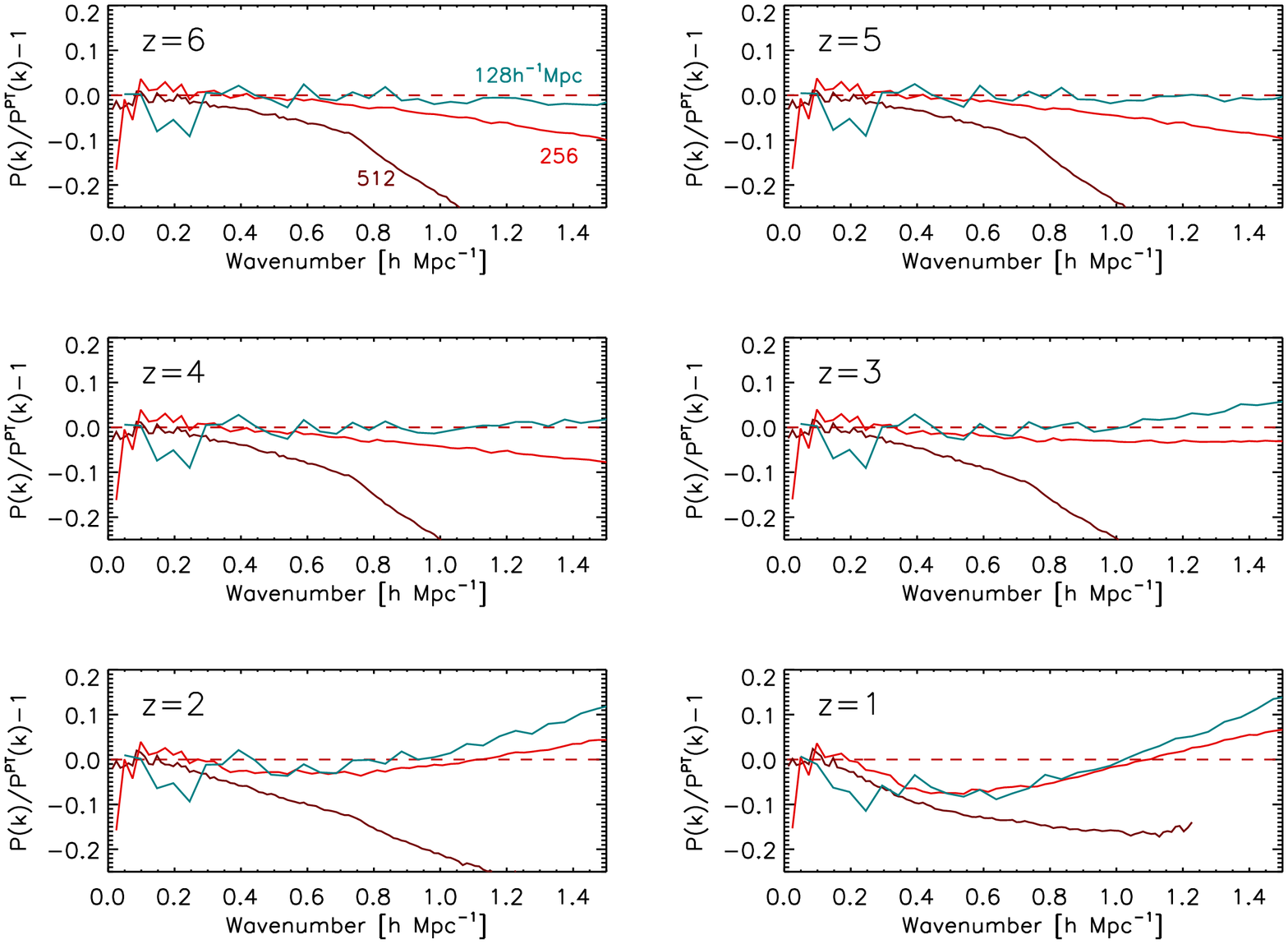}
}%
\rotatebox{90}{%
  \includegraphics[width=6.5cm]{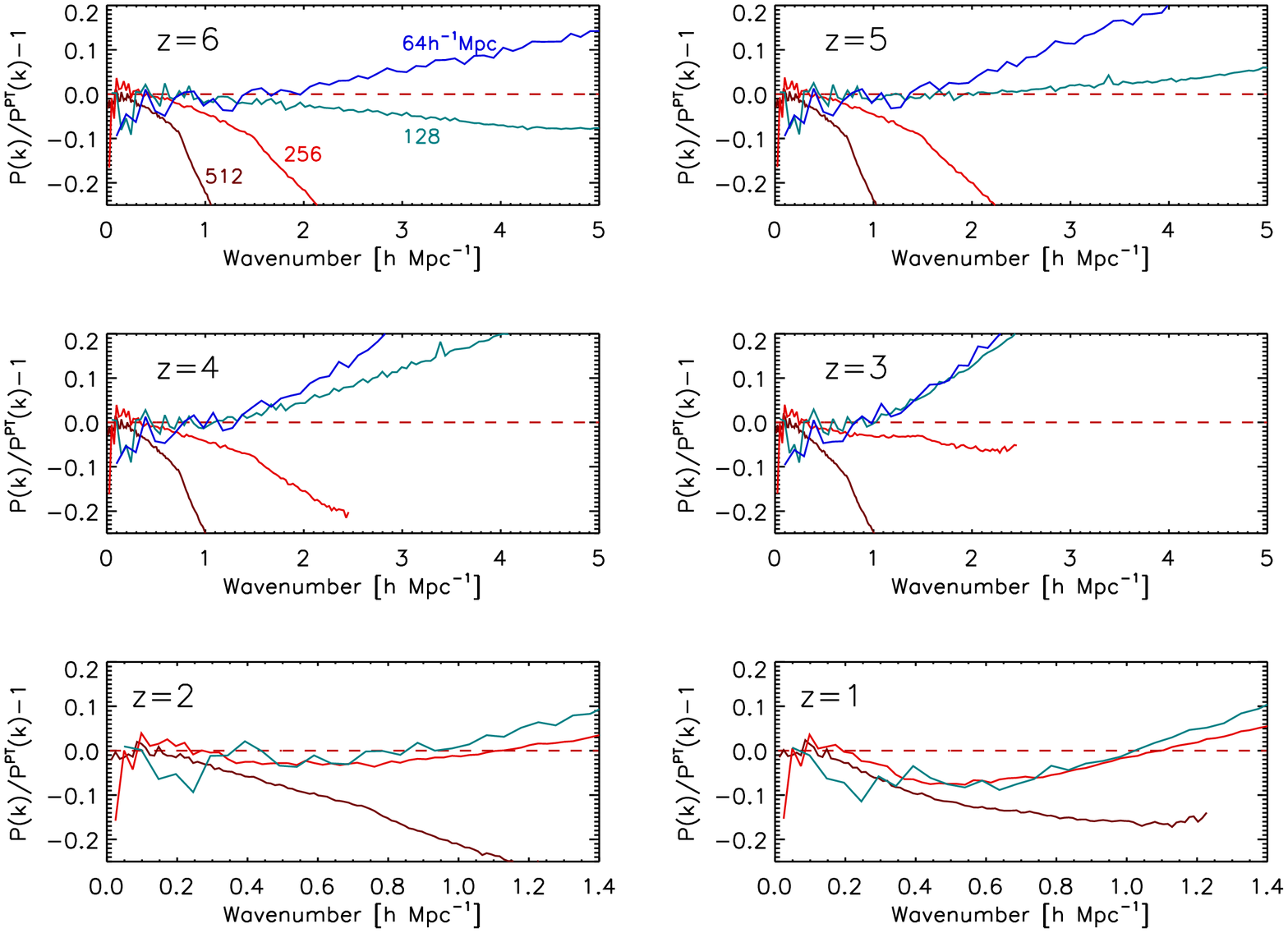}
}%
\caption{%
Convergence test.
({\it Left}) Fractional differences between the power spectra from
 $N$-body simulations in $L_{\rm box}=512$, 256, and 128~$h^{-1}$~Mpc
 box (from bottom to top lines) and the perturbation theory predictions
 in $k<1.5~h~{\rm Mpc}^{-1}$. 
({\it Right}) The same as left panel, but for simulations in
$L_{\rm box}=512$, 256, 128, and 64~$h^{-1}$~Mpc
 box (from bottom to top lines) in the expanded range in wavenumber,
$k<5~h~{\rm Mpc}^{-1}$. 
}%
\label{fig5}
\end{figure}

Figure~\ref{fig5} shows that simulations with a larger box size
lack power on larger scales due to the lack of resolution, as expected, 
while they have better statistics on large scales than those with a 
smaller box size. This figure helps us to determine the valid range
in wavenumber from each simulation box. 
We find that one can use 512, 256, 128, and 64~$h^{-1}$~Mpc simulations
to calculate reliable estimates of the power spectrum
in $k\leq 0.24~h~{\rm Mpc}^{-1}$, $0.24<k\leq 0.5~h~{\rm Mpc}^{-1}$,
$0.5<k\leq 1.4~h~{\rm Mpc}^{-1}$, and
$1.4<k\leq 5~h~{\rm Mpc}^{-1}$, respectively.

\end{document}